\documentclass[12pt]{article}
\usepackage{graphicx}
\usepackage{epstopdf}
\usepackage{caption}
\usepackage{subcaption}
\usepackage{subfig}

\usepackage{epsfig,amsmath,amssymb,verbatim,mathrsfs}

\def\beq{\begin{eqnarray}}
\def\eeq{\end{eqnarray}}
\def\bea{\begin{eqnarray}}
\def\eea{\end{eqnarray}}

\begin{document}

\setlength{\baselineskip}{0.2in}


\begin{titlepage}
\noindent
\flushright{March 2013}
\vspace{0.2cm}

\begin{center}
  \begin{Large}
    \begin{bf}
Generalized Inverse Seesaws\\

     \end{bf}
  \end{Large}
\end{center}

\vspace{0.2cm}

\begin{center}

\begin{large}

{Sandy~S.~C.~Law$^{*}$\footnote{slaw@mail.ncku.edu.tw}  and Kristian~L.~McDonald$^{\dagger}$\footnote{klmcd@physics.usyd.edu.au}}\\
     \end{large}
\vspace{0.5cm}
  \begin{it}
* Department of Physics, National Cheng-Kung University, \\ Tainan 701, Taiwan\\
\vspace{0.5cm}
$\dagger$ ARC Centre of Excellence for Particle Physics at the Terascale,\\
School of Physics, The University of Sydney, NSW 2006, Australia\\\vspace{0.5cm}
\vspace{0.3cm}
\end{it}
\vspace{0.5cm}

\end{center}


\begin{abstract}

The seesaw mechanism can be generalized to a Type-III variant and a quintuplet variant. We present two models that provide analogous generalizations of the inverse seesaw mechanism. The first model employs a real fermion triplet $\mathcal{F}\sim(1,3,0)$ and  requires no additional multiplets or parameters relative to the standard inverse seesaw. We argue that, from a bottom-up perspective, there appears to be no particular reason to preference the usual scenario over this variant. The second model employs a fermion quintuplet $\mathcal{F}\sim(1,5,0)$ and requires an additional scalar $S\sim(1,4,1)$. We also show that minimal inverse seesaws with even larger fermionic representations are not expected to realize naturally small neutrino masses.

\end{abstract}

\vspace{1cm}

\end{titlepage}

\setcounter{page}{1}


\vfill\eject


\section{Introduction\label{sec:introduction}}

It is well known that the canonical (or Type-I) seesaw mechanism~\cite{type1_seesaw} can be generalized to a Type-III variant~\cite{Foot:1988aq}. Recently Ref.~\cite{Kumericki:2012bh} has shown that a quintuplet variant also exists. Actually, these are the only three possibilities for a minimal tree-level seesaw due to heavy fermion exchange when lepton-number symmetry is broken by a single (Majorana) mass insertion~\cite{McDonald:2013kca}. 

The inverse seesaw mechanism, on the other hand,  provides an interesting alternative to the conventional seesaws~\cite{inverse_seesaw}. This approach requires an increased field content, but has the advantage of lowering the mass scale for the new physics, making it eminently more testable. The majority of works that have thus far studied the inverse seesaw have employed what could be called a standard, or Type-I, inverse seesaw. In this work we show that the inverse seesaw mechanism can be generalized to a Type-III variant, and a quintuplet variant, in complete analogy with the conventional seesaw. These results fill some gaps in the literature and further map the ``theory space" for neutrino mass models.

As we shall see, the Type-III inverse seesaw employs a real fermion triplet and turns out to be a very simple generalization. It does not require any additional parameters or multiplets, relative to the standard inverse seesaw, and yet retains the appealing features of the standard approach. As best we can tell, there appears to be no reason to preference the standard inverse seesaw over this variant.

The quintuplet inverse seesaw requires an additional scalar but otherwise operates much the same as the conventional version. It has some interesting phenomenological differences, however. Beyond these two generalizations, we find that further minimal variants with larger fermion multiplets are not as interesting due to the absence of a natural explanation for a requisite small vacuum expectation value (VEV).

Before proceeding we note that the small lepton number violating masses found in inverse seesaws can be radiatively induced in the so called ``radiative inverse seesaws"~\cite{Ma:2009gu}. The LHC phenomenology of the triplet fermion employed in the Type-III seesaw is detailed in Ref.~\cite{Franceschini:2008pz,delAguila:2008hw}. For recent works relating neutrino masses to the existence of larger fermion representations see Refs.~\cite{Kumericki:2012bh,McDonald:2013kca,Babu:2009aq,Picek:2009is,Ren:2011mh}, while some bounds on large scalar multiplets are detailed in Refs.~\cite{Hally:2012pu,Earl:2013jsa} (also see Ref.~\cite{Liao:2010rx}). Also note that  for simplicity we present our discussion for a single generation, though all results are easily generalized.

The plan of this paper is as follows. In Section~\ref{sec:inverse_seesaw} we provide a brief overview of the inverse seesaw mechanism, reminding readers of features relevant to our subsequent discussion.  Section~\ref{sec:general_inverse_seesaw} describes the basic logic that guides our generalization process, and presents the simplest (Type-III) generalization. We move beyond the simplest generalization in Section~\ref{sec:non-minimal_inverse_seesaw} to arrive at a quintuplet inverse seesaw. Conclusions are drawn in Section~\ref{sec:conc}.

\section{The Inverse Seesaw Mechanism\label{sec:inverse_seesaw}}
The SM can be extended to realize an inverse seesaw mechanism by adding a vector-like gauge-singlet fermion, $N=N_R+N_L\sim(1,1,0)$, to the particle spectrum. The Lagrangian then contains the terms
\bea
\mathcal{L}&\supset&  i\bar{N}\gamma^\mu\partial_\mu N - M\bar{N} N -\frac{\delta_R}{2}\,\overline{N_R^c}N_R  -\frac{\delta_L}{2}\,\overline{N_L^c}N_L  -\lambda\bar{L} \tilde{H} N_R+\mathrm{H.c.}.
\eea
Here $L$ ($H$) is the SM lepton (scalar) doublet, and the tilde denotes charge conjugation. Gauge invariance also permits a term $\overline{L}\tilde{H} N_L^c$, however a field redefinition can be performed to rotate this term away. One can therefore set this term to zero without loss of generality. Lepton number symmetry is explicitly broken by the bare Majorana masses $\delta_{L,R}$, which are consistent with gauge invariance.

After electroweak symmetry breaking the SM neutrino $\nu_L$ mixes with the singlet fermion. In the basis  $\{\nu_L,N_R^c,N_L\}$, the full neutral lepton mass matrix is given by
\bea
\mathcal{M}_{\mathrm{inv}}=
\left(
\begin{array}{ccc}
0 & m_D& 0\\
m_D &\delta_R& M\\
0&M&\delta_L
\end{array}
\right),\label{eq:inv_seesaw_mass_matrix}
\eea
where the Dirac mass is $m_D=\lambda\langle H\rangle$. The inverse seesaw mechanism occurs in the technically-natural limit of $\delta_{L,R}\ll m_D,\,M$, for which the eigenvalues are
\bea
 m_\nu &=& \frac{m_D^2\delta_L}{m_D^2+M^2} + \mathcal{O}(\delta^2) \,, \label{eqn:evals_1}\\
 m_{2,3} &=&
 \mp \sqrt{m_D^2+M^2} + \frac{M^2\delta_L}{2(m_D^2+M^2)} +\frac{\delta_R}{2}
 + \mathcal{O}(\delta^2)\,.\label{eqn:evals_23}
\eea
Observe that the light neutrino mass (i.e. $m_\nu$) is suppressed due to the smallness of $\delta_L$, with further suppression occurring for $m_D \ll M$. 

An interesting feature of the inverse seesaw is that it permits significant mixing between $\nu_L$ and $N_L$, namely $\theta \simeq \mathcal{O}(m_D/M)$. We can write this as
\bea
\theta &\sim& \frac{m_D}{M} \ \sim\ 10^{-2} \times \left(\frac{\mathrm{TeV}}{M}\right)\times \left(\frac{\lambda}{0.1}\right).
\eea
Thus, for $M\sim$~TeV and $\lambda\sim0.1$  the mixing is $\mathcal{O}(10^{-2})$ and non-unitary mixing effects may be observable in future experiments~\cite{non_unitary_exp}. Of course, the light neutrino mass is naturally small even if these new effects are not within experimental reach, though new observable phenomenology would certainly be welcome. Also note that the mixing between $\nu_L$ and $N_R$ is at the level of $\mathcal{O}(m_\nu/m_D) \simeq \mathcal{O}(10^{-9})$ and remains negligible.\footnote{For these parameters, a normal type-I/III seesaw gives mixing of $\mathcal{O}(10^{-7})$.}
\section{Generalizing the Inverse Seesaw\label{sec:general_inverse_seesaw}}
\begin{figure}[ttt]
\begin{center}
        \includegraphics[width = 0.6\textwidth]{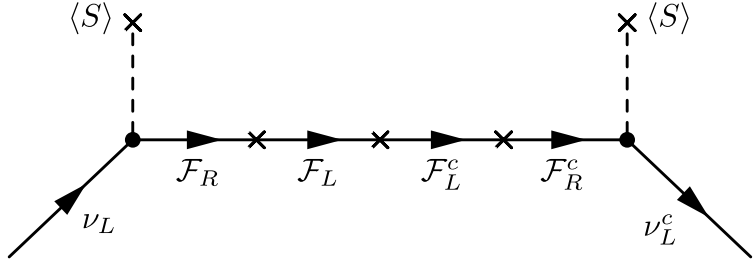}
\end{center}
\caption{Tree-level diagram for a generalized inverse seesaw mechanism. The simplest realizations occur when the scalar is the SM doublet, $S=H\sim(1,2,1)$, and the vector-like fermion $\mathcal{F}\sim(1,R_{\mathcal{F}},0)$ is either a gauge singlet ($R_{\mathcal{F}}=1$) or a triplet ($R_{\mathcal{F}}=3$).  The former case is the standard inverse seesaw mechanism but the latter (Type-III) case is equally simple.}\label{fig:general_inverse}
\end{figure}

The inverse seesaw mechanism described in the previous section can be reduced to the following (generalized) elements:
\begin{itemize}
\item The SM is extended to include a real vector-like  fermion: $\mathcal{F}=\mathcal{F}_R+\mathcal{F}_L\sim (1,R_{\mathcal{F}},0)$. Here $R_{\mathcal{F}}$ must be odd to ensure $\mathcal{F}$ contains a neutral component. 
\item The new fermion Yukawa-couples to the SM lepton doublet, $\mathcal{L}\supset -\lambda \overline{L} \tilde{S} \mathcal{F}_R$, where the scalar $S$ transforms as $S\sim(1,R_S,1)$. 
\item The quantum numbers of the scalar should satisfy $R_S\otimes R_L\supset R_{\mathcal{F}}$, where $R_L=2$.
\item Lepton number symmetry is broken explicitly by bare Majorana mass terms for the chiral components of $\mathcal{F}$.
\end{itemize} 
Reducing the inverse seesaw to these basic ingredients makes it apparent that simple generalizations are possible. Indeed, a generic tree-level diagram for a generalized inverse seesaw can be drawn; see Figure~\ref{fig:general_inverse}. Minimal realizations of the inverse seesaw occur when $S=H\sim(1,2,1)$ is the SM scalar, as no  new scalars are required. However, the fermion $\mathcal{F}$ is a beyond-SM field and the use of the gauge-singlet field $\mathcal{F}\sim(1,1,0)$  is not the only possibility, as we now show.

 The simplest generalization of the inverse seesaw arises when one retains $S=H$ but employs a fermion with $R_{\mathcal{F}}>1$. The only such candidate compatible with gauge invariance is the real fermion triplet $\mathcal{F}\sim(1,3,0)$. In fact, this is a particularly simple generalization that retains all the appealing features of the (standard) inverse seesaw. The Lagrangian contains the terms 
\bea
\mathcal{L}&\supset&  i\bar{\mathcal{F}}\gamma^\mu D_\mu \mathcal{F}- M\bar{\mathcal{F}} \mathcal{F} -\frac{\delta_R}{2}\,\overline{\mathcal{F}_R^c}\mathcal{F}_R  -\frac{\delta_L}{2}\,\overline{\mathcal{F}_L^c}\mathcal{F}_L  -\lambda\bar{L} \tilde{H} \mathcal{F}_R+\mathrm{H.c.},\label{eq:type_III_lagrangian}
\eea
and once again a field redefinition is performed, without loss of generality, to remove the term $\overline{L} \tilde{H} \mathcal{F}_L^c$. After electroweak symmetry breaking there is mass mixing between the SM neutrino $\nu_L$ and the new neutral fermions. We again label this Dirac mass as $m_D=\lambda \langle H\rangle$, and write the neutral-fermion mass matrix as
\bea
\mathcal{M}_{\mathrm{III}}=
\left(
\begin{array}{ccc}
0 & m_D& 0\\
m_D &\delta_R& M\\
0&M&\delta_L
\end{array}
\right),\label{eq:type_3_mass_matrix}
\eea
in the basis  $\{\nu_L,\mathcal{F}_R^c,\mathcal{F}_L\}$. This has the same form as the inverse seesaw mass matrix in Eq.~\eqref{eq:inv_seesaw_mass_matrix}. Thus, it is apparent that the mass eigenvalues have the same form as Eqs.~\eqref{eqn:evals_1} and~\eqref{eqn:evals_23}.

This model provides a technically natural explanation for the small neutrino masses observed in Nature, on par with the standard inverse seesaw. Note that this generalization is no more complicated than the standard realization; relative to the inverse seesaw no additional particle multiplets are required and the number of new parameters is the same. From a bottom-up perspective we see no particular reason to favor the minimal (or Type-I) inverse seesaw relative to this triplet version. In analogy with the Type-III seesaw~\cite{Foot:1988aq},  one can refer to this generalization as a Type-III inverse seesaw. Also note that, as with the standard incarnation, when the new fermion is at the TeV scale ($M\sim$~TeV) one can obtain observable non-unitary mixing effects.

We note that related inverse seesaws, with triplet leptons, have appeared in the literature. Ref.~\cite{Abada:2007ux}, in which  low-energy phenomenological effects from effective operators with dimension $d=6$ (related to the origin of neutrino mass) are rigorously studied, also mentions triplet leptons in relation to the inverse seesaw. Refs.~\cite{Ma:2009kh,JosseMichaux:2012wj,Morisi:2012hu} also employ hypercharge-less triplet leptons and realize neutrino mass via non-standard inverse seesaws. These works differ from the approach discussed here, however,  as they all contain beyond-SM symmetries, under which the triplets have non-trivial charges. Thus, the set of bare triplet-mass terms shown in Eq.~\eqref{eq:type_III_lagrangian} [and employed in Eq.~\eqref{eq:type_3_mass_matrix}] are not allowed in these works, and an extended field content is required to generate neutrino mass via an inverse seesaw. 

Triplet leptons were employed in Ref.~\cite{delAguila:2008hw}, though the mass matrix used there differs from the most general one shown in Eq.~\eqref{eq:type_3_mass_matrix}, as they ignore the (allowed)  Majorana mass $\delta_R$. Ref.~\cite{delAguila:2008hw} does, however, consider the collider phenomenology of the heavy (pseudo-) Dirac neutral-fermions contained in the vector-like triplet. Interestingly they find that the neutral fermion is readily distinguished from its Majorana counterpart in the Type-III seesaw. In both cases, final states with  as many as six leptons are discussed. They find that, in the Dirac case, like-sign dilepton signals ($\ell^\pm\ell^\pm$) are less significant, while three-lepton signals ($\ell^\pm\ell^\pm\ell^\mp$) and four-lepton signals ($\ell^+\ell^+\ell^-\ell^-$) have a sizable (and similar) significance. This differs from the Majorana case where the four-lepton signal is smaller while the two- and three-lepton signals have a similar significance. The vector-like triplet leptons can therefore be readily discriminated from their Majorana counterpart by neutral-fermion production at the LHC.

Before moving on to consider other fermionic representations, we note that one can also use the scalar $S\sim(1,4,1)$ in conjunction with the triplet fermions to realize an inverse seesaw. This scalar has an allowed renormalizable coupling of the form $\overline{F}_R SL$, and can therefore play the role of the external legs shown in Figure~\ref{fig:general_inverse}. However,  the doublet contribution will also be present, so this would only produce a modification of the Type-III inverse seesaw already described; viable neutrino masses would not depend on the presence or absence of the quadruplet scalar. Similarly, if one considers an extra SM-like doublet, this would not be needed to generate neutrino mass, and the resulting model would simply be a more cumbersome variant of the minimal scenario described here.
\section{Beyond the Minimal Inverse Seesaws\label{sec:non-minimal_inverse_seesaw}}
The minimal inverse seesaws, described above, arise when one employs a fermion $\mathcal{F}\sim(1,R_{\mathcal{F}},0)$ with $R_{\mathcal{F}}=1$ or $R_{\mathcal{F}}=3$. However, one can move beyond these minimal implementations and ask if further variations are possible for $R_{\mathcal{F}}>3$. The next simplest choice is $R_{\mathcal{F}}=5$, giving a fermion quintuplet $\mathcal{F}\sim(1,5,0)$. In this case the SM scalar is no longer sufficient to generate the Yukawa coupling between the lepton doublet $L$ and the new fermion. The minimal choice for a new scalar is $S\sim(1,4,1)$, which gives the new Lagrangian terms
\bea
\mathcal{L}&\supset&  i\bar{\mathcal{F}}\gamma^\mu D_\mu \mathcal{F}- M\bar{\mathcal{F}} \mathcal{F} -\frac{\delta_R}{2}\,\overline{\mathcal{F}_R^c}\mathcal{F}_R  -\frac{\delta_L}{2}\,\overline{\mathcal{F}_L^c}\mathcal{F}_L  -\lambda\bar{L} \tilde{S} \mathcal{F}_R+\lambda_S \tilde{S} \tilde{H} H^\dagger H.
\eea
Note the new quartic term in the scalar potential. This term, linear in $S$, forces $S$ to develop a nonzero VEV after the SM scalar triggers electroweak symmetry breaking:\footnote{This is analogous to the VEV suppression found in the Type-II seesaw~\cite{type2_seesaw}.}
\bea
\langle S\rangle& \simeq &\lambda_S \,\frac{\langle H\rangle^3}{M_S^2},\label{eq:new_vev}
\eea
where $M_S$ is the mass term for $S$. Observe that this VEV is naturally suppressed, relative to the weak scale, for $M_S\gg \langle H\rangle$. This feature is important as the new scalar contributes to electroweak symmetry breaking and to the SM gauge boson masses. In particular, it modifies the tree-level value of the $\rho$-parameter away from the SM prediction of  $\rho=1$. Experimental constraints require $\langle S\rangle \lesssim\mathcal{O}(1)$~GeV~\cite{Nakamura:2010zzi} which is readily obtained, without fine tuning,  due to the inverse dependence of the scalar mass in Eq.~\eqref{eq:new_vev}.

Beyond these initial differences, this model is actually very similar to a standard inverse seesaw. The mass matrix again has the form of Eq.~\eqref{eq:type_3_mass_matrix}, but now the Dirac mass coupling the SM neutrino to the new fermion is given by $m_D=\lambda\langle S\rangle$. For ease of exposition, let us redisplay the mass eigenvalues
\bea
 m_\nu &=& \frac{m_D^2\delta_L}{m_D^2+M^2} + \mathcal{O}(\delta^2) \quad \sim \quad\frac{\lambda^2\langle S\rangle^2}{M^2}\,\delta_L\,, \label{eqn:evals_1_nonminimal}\\
 m_{2,3} &=&
 \mp \sqrt{m_D^2+M^2} + \mathcal{O}(\delta)\quad\sim \quad\mp M\,,\label{eqn:evals_23_nonminimal}
\eea
which have the same form as the minimal cases of $R_{\mathcal{F}}=1$ and $R_{\mathcal{F}}=3$, modulo the replacement $\langle H\rangle\rightarrow \langle S\rangle$. This difference has an important consequence; 
noting that the mixing between $\nu_L$ and $\mathcal{F}_L$ is $\theta\sim \mathcal{O}(m_D/M)$, one obtains
\bea
\theta &\sim& \frac{m_D}{M} \ \sim\ 10^{-2} \times \left(\frac{\mathrm{TeV}}{M}\right)\times \left(\frac{\lambda}{0.1}\right)\times \left(\frac{\langle S\rangle}{\langle H\rangle}\right).
\eea
Thus, for given fixed values of $M$ and $\lambda$, the mixing in the quintuplet model is suppressed by a factor of $\langle S\rangle/\langle H\rangle\ll1$ relative to the triplet model. However,  we note that for $M\sim$~TeV and $\langle S\rangle\sim$~GeV this mixing is at the level of $\theta\sim 10^{-3}$ for $\lambda\sim\mathcal{O}(1)$, which is more difficult to experimentally probe, though not beyond conceivable experimental reach. Certainly, non-unitary mixing effects remain much easier to observe in the quintuplet model compared to the standard Type-I and Type-III seesaws. The triplet and quintuplet models also contain additional charged fermions that enable one to differentiate between the models.

Note that an interesting seesaw model employing a chiral fermion quintuplet, $\mathcal{F}_R\sim(1,5,0)$, and the  beyond-SM scalar $S\sim(1,4,1)$, was introduced in Ref.~\cite{Kumericki:2012bh} (also see Ref.~\cite{McDonald:2013kca}). Despite the similar particle content, the model of Ref.~\cite{Kumericki:2012bh} is very different from the present model. Ref~\cite{Kumericki:2012bh} achieves neutrino masses with the (generalized) seesaw form $m_\nu\sim\langle S\rangle^2/M$,  whereas we obtain the generalized inverse seesaw form of $m_\nu\sim (\langle S\rangle/M)^2\times\delta_L$. Actually, this difference is easy to explain; the relationship between the model of Ref.~\cite{Kumericki:2012bh} and the present model is completely analogous to the difference between a Type-I seesaw~\cite{type1_seesaw}, in which the SM is extended by the addition of a chiral gauge-singlet field, and the standard inverse seesaw, in which a vector-like singlet is introduced. In the former case the light neutrino mass has the form $m_\nu\sim \langle H\rangle^2/M$, while in the inverse seesaw one has $m_\nu \sim (\langle H\rangle /M)^2\times\delta_L$, the difference between which is completely analogous to that of the two  quintuplet fermion models.\footnote{If one was to refer to the model of Ref.~\cite{Kumericki:2012bh}  as a ``Type-V seesaw," the quintuplet model  presented here would be a ``Type-V inverse seesaw." This labeling correlates the size of the fermion representation employed in the seesaw with the number for the seesaw type; a seesaw with the fermion $\mathcal{F}\sim(1,R_{\mathcal{F}},0)$ is a Type-$R_{\mathcal{F}}$ seesaw.} This relationship between neutrino masses in the standard seesaws and the analogous inverse seesaws is summarized in Table~\ref{seesaws}, with Yukawa couplings set to unity for simplicity. 

Ref.~\cite{Kumericki:2012bh} considered some phenomenological aspects of the Majorana quintuplet fermion employed in the quintuplet seesaw. As we noted in the triplet-fermion case, the difference between the  heavy Majorana-fermion in a Type-III seesaw and the heavy pseudo-Dirac fermion in the Type-III inverse seesaw is experimentally discernible~\cite{delAguila:2008hw}. The different signals in these cases can be traced back to the Majorana vs. Dirac nature of the neutral fermion. This difference also exists between the heavy neutral fermion in the quintuplet seesaw of Ref.~\cite{Kumericki:2012bh} and the inverse variant presented here; in the inverse seesaw the heavy fermion is Dirac-like and thus some final states containing leptons will have a different relative significance compared to the Majorana case. Of course, the doubly-charged fermions in the quintuplet models will likely give more pronounced signals, but the different signals from neutral-fermion production can help discriminate the models.


\begin{table}
\centering
\begin{tabular}{|c|c|c|c|c|}\hline
& \multicolumn{2}{c|}{}&\multicolumn{2}{c|}{} \\
&
\multicolumn{2}{c}{Seesaw}& \multicolumn{2}{|c|}{Inverse Seesaw} \\
& \multicolumn{2}{c|}{}&\multicolumn{2}{c|}{} \\\cline{2-5}
Model&&&&\\
& New Multiplets& $m_\nu$& New Multiplets& $m_\nu$\\
&&&&\\
\hline
&&&&\\
$\quad$Type-I$\quad$ & $\ N_R\sim(1,1,0)\ $ &$\quad\langle H\rangle^2/M\quad$ & $\ N\sim(1,1,0)\ $& $\quad(\langle H\rangle/M)^2\times \delta_L\quad$\\
&&&&\\
\hline
&&&&\\
$\quad$Type-III$\quad$ & $\ \mathcal{F}_R\sim(1,3,0)\ $ &$\quad\langle H\rangle^2/M\quad$ & $\ \mathcal{F}\sim(1,3,0)\ $& $\quad(\langle H\rangle/M)^2\times \delta_L\quad$\\
&&&&\\
\hline
&&&&\\
Quintuplet & $\ \mathcal{F}_R\sim(1,5,0)\ $ & & $\ \mathcal{F}\sim(1,5,0)\ $& \\
or& &$\quad\langle S\rangle^2/M\quad$ &&$\quad(\langle S\rangle/M)^2\times \delta_L\quad$\\
Type-V&$S\sim(1,4,1)$&&$S\sim(1,4,1)$&\\
&&&&\\
\hline
\end{tabular}
\caption{\label{seesaws} Conventional Seesaws and the Corresponding Inverse Seesaws.}
\end{table}

In the context of the present work it is natural to ask if further generalizations of the inverse seesaw are possible. Moving beyond the quintuplet with $R_{\mathcal{F}}=5$, the next possibility is $\mathcal{F}\sim(1,7,0)$. In this case the SM must also be extended to include a new scalar with the quantum numbers $S\sim(1,6,1)$ or $S\sim(1,8,1)$. While one can consider such an extension, and realize an inverse seesaw mass of the form $m_\nu\sim(\langle S\rangle/M)^2\times \delta_L$, this model is less desirable. As in the quintuplet model, the VEV of the new scalar is bound by $\rho$-parameter measurements to roughly obey  $\langle S\rangle \lesssim\mathcal{O}(1)$~GeV. However, unlike the quintuplet model, the quantum numbers of the candidate new scalars do not permit a term linear in $S$ in the scalar potential $V(H,S)$. Therefore the new scalar does not develop an induced VEV after electroweak symmetry breaking; one must instead engineer the relation $\langle S\rangle\ll M_S$ by tuning parameters in the scalar potential. Small neutrino masses therefore require some tuning in this case, if they are to be consistent with electroweak precision constraints; if we restrict our attention to natural generalizations of the inverse seesaw  we can disregard this model.\footnote{To be clear, minimal models with $R_{\mathcal{F}}>5$ do not allow a generalized version of the VEV suppression found in the Type-II seesaw.}

The above statements also hold for even larger fermion multiplets ($R_{\mathcal{F}}>7$). The requisite scalar multiplets remain too large to allow a naturally small VEV, and some tuning is necessary to achieve light SM neutrino masses and satisfy electroweak precision constraints. Furthermore, the above considerations also apply if one employs the quintuplet fermion in concert with $S\sim(1,6,1)$. This is the reason we did not discuss this possibility above, and instead restricted our attention to the case with $S\sim(1,4,1)$ when $\mathcal{F}\sim(1,5,0)$.

Finally, let us also point out that Ref.~\cite{Ibanez:2009du} has considered an inverse seesaw with a chiral fermion triplet $\mathcal{F}_R\sim(1,3,0)$. Their approach differs from the minimal Type-III inverse seesaw presented in Section~\ref{sec:general_inverse_seesaw}; they replace $N_R\rightarrow \mathcal{F}_R$ but retain $N_L$ and thus require the additional scalar $S\sim(1,3,0)$ to achieve a Dirac coupling between $\mathcal{F}_R$ and $N_L$. This differs from our Type-III model with the vector-like triplet, in which no additional parameters or multiplets are required relative to the standard inverse seesaw. It should be apparent from our work that a number of generalizations are possible once one admits distinct chiral fermion multiplets, as in Ref.~\cite{Ibanez:2009du}. For example, one could replace $N_R\rightarrow \mathcal{F}_R\sim(1,5,0)$, along with the requisite scalars, and retain $N_L$. Alternatively one can generalize the inverse seesaw models so that two distinct external scalar legs appear in Figure~\ref{fig:general_inverse}, as was recently done for the Type-III seesaw in Ref.~\cite{Ren:2011mh}. These generalizations are, however, less minimal than the triplet and quintuplet inverse seesaws discussed here, in the sense that additional field-multiplets are required.





\section{Conclusion\label{sec:conc}}
We have studied generalized inverse seesaws in which the vector-like gauge-singlet neutrino of the standard inverse seesaw is replaced by a real vector-like  fermion $\mathcal{F}\sim(1,R_{\mathcal{F}},0)$, with odd-valued $R_{\mathcal{F}}$. Two interesting generalizations were found. The (Type-III) model with $R_{\mathcal{F}}=3$ provides a particularly simple generalization as it does not require any additional parameters or multiplets relative to the standard inverse seesaw. Indeed, from a bottom-up perspective there does not appear to be any reason to favor the standard inverse seesaw over this variant. The quintuplet model with $R_{\mathcal{F}}=5$ also realizes a viable inverse seesaw, though non-unitary mixing effects are more difficult to observe in this variant. In minimal inverse seesaws with larger values of $R_{\mathcal{F}}>5$ one requires some tuning to obtain small neutrino masses and remain consistent with electroweak precision constraints, making such cases less interesting.
\section*{Acknowledgments\label{sec:ackn}}
SSCL is supported in part by the NSC under Grant No.
NSC-101-2811-M-006-015 and in part by the NCTS of Taiwan. KM is supported by the Australian Research Council.

\end{document}